\documentclass[11pt]{article}
\usepackage{latexsym}
\usepackage{amssymb}
\usepackage{epsf}
\usepackage{amsmath}
\usepackage[hypertex]{hyperref}
\usepackage{graphicx}

\begin{document}

\pagestyle{empty}

\begin{flushright}
LA-UR-08-0384\\
\end{flushright}

\vspace{2.5cm}

\begin{center}

{\bf\LARGE A Clean Slepton Mixing Signal at the LHC}
\\

\vspace*{1.5cm}
{\large 
Ryuichiro Kitano
} \\
\vspace*{0.5cm}

{\it Theoretical Division T-8, Los Alamos National Laboratory, Los Alamos, NM 87545}\\
\vspace*{0.5cm}

\end{center}

\vspace*{1.0cm}

\begin{abstract}
{\normalsize
\baselineskip 14pt
In supersymmetric scenarios where the scalar tau lepton is stable or
 long-lived, a search for a decay mode $\chi^0 \to \tilde \tau \mu$ at
 the LHC has a good sensitivity to the flavor mixing in the scalar
 lepton sector. We demonstrate that the sensitivities to the mixing
 angle at the level of $\sin \theta_{23} = 0.15$ are possible with an
 integrated luminosity of 100~fb$^{-1}$ if the total production cross
 section of supersymmetric particles is of the order of 1~pb.  The
 sensitivity to the mixing parameter can be better than the experimental
 bound from the $\tau \to \mu \gamma$ decay depending on model
 parameters.

}
\end{abstract} 

\newpage
\baselineskip 18pt
\setcounter{page}{2}
\pagestyle{plain}

\setcounter{footnote}{0}

If new physics contains a charged stable particle, such as the scalar
tau lepton ($\tilde \tau$) in supersymmetric (SUSY) models, it provides
a very clean signal at the LHC experiments. Once they are produced, most
of them penetrate detectors and leave charged tracks just like muons. By
measuring the velocity at the muon system, we can easily distinguish
from the muon background. A very precise mass measurement is possible by
combining with the momentum measurements~\cite{Ambrosanio:2000ik}.

Scenarios with such a charged stable or long-lived particle have sounded
exotic and regarded as alternative possibilities. However, recent
serious considerations of SUSY phenomenology have shown that it is
indeed theoretically motivated~\cite{Ibe:2007km}.
The presence of such particles does not immediately contradict with
cosmological history. There are interesting cosmological scenarios and
even motivations for such a long-lived
particle~\cite{Feng:2004zu,Cembranos:2005us,Pospelov:2006sc,Kohri:2006cn}.
If it is the case, we will have new kinds of signals in new physics
search experiments.

In this paper, we propose a search strategy for flavor mixing in the
scalar lepton sector in the stable (or long-lived) $\tilde \tau$
scenario at the LHC. In the presence of the flavor mixing, we will have
a decay mode of the neutralinos $\chi^0 \to \tilde \tau \mu$. By looking
for sharp peaks in the $\tilde \tau$-$\mu$ invariant mass, we show that
we will be able to discover lepton flavor violation for $\Gamma(\chi_1^0
\to \tilde \tau \mu)/\Gamma(\chi_1^0 \to \tilde \tau \tau) \simeq \tan^2
\theta_{23} \gtrsim 10^{-2}$, where $\theta_{23}$ is the slepton mixing
angle.

There have been many studies on lepton flavor violation at the LHC and
$e^+ e^-$ colliders assuming the neutralino to be the lightest SUSY
particle. The possibility of observing $e^\pm \mu^\mp $ + missing $E_T$
final states at $e^+ e^-$ colliders has been pointed out in
Ref.~\cite{Krasnikov:1994hr}.
The correct treatment of the process including quantum interference
(slepton oscillation) has been studied in Ref.~\cite{ArkaniHamed:1996au}
and discussion has been expanded to the LHC experiments and CP violation
in Ref.~\cite{ArkaniHamed:1997km}. Following those papers, LHC studies
on searches for decay processes $\chi^0_2 \to l_i^\pm l_j^\mp \chi_1^0$
with $i \neq j$ have been done in Refs.~\cite{Agashe:1999bm}. The
sensitivities of $O(0.1)$ for mixing angles have been derived in various
SUSY models.

Lepton flavor violation in the long-lived $\tilde \tau$ scenario has
also been studied. In Ref.~\cite{Hamaguchi:2004ne}, the decay of $\tilde
\tau$ into $e$ or $\mu$ and a gravitino is studied under an assumption
that a significant number of $\tilde \tau$'s will be collected at the
LHC or future linear collider experiments by placing a massive stopper
material close to the detectors~\cite{Hamaguchi:2004df}.
A linear collider study with long-lived $\tilde \tau$ has also been done
in Ref.~\cite{Ibarra:2006sz} where it is proposed to search for lepton
flavor violating final states such as $(e^+ \tau^\pm \tilde \tau^{\mp})
\tilde \tau^-$ through slepton pair production processes.
Very good sensitivities as well as $\sin \theta \sim$ (a
few)$\times 10^{-2}$ are reported in both of the works.
We study in the following the LHC signals of slepton flavor mixing
without new detectors or future colliders. Therefore it serves as the
first search strategy that can be done immediately after the LHC starts
if $\tilde \tau$ is stable or long-lived.

In order to estimate the sensitivity, we performed the following Monte
Carlo simulation. We used a model of Ref.~\cite{Ibe:2007km} where the spectrum of
the SUSY particles are parametrized by four quantities:
\begin{eqnarray}
 \mu,\ \ \ M_{\rm gaugino} (\equiv M_{\tilde g} / g_3^2),\ \ \ M_{\rm
  mess}, \ \ \ N_{\rm mess}.
\end{eqnarray}
The $\mu$ parameter and $M_{\rm gaugino}$ control the Higgsino mass and
the gaugino masses, respectively. The messenger scale $M_{\rm mess}$ and
the number of messenger particle $N_{\rm mess}$ determines masses of
scalar particles relative to the gaugino masses. We have chosen two
parameter points where $\tilde \tau$ is the lightest SUSY particle
(except for the gravitino):
\begin{eqnarray}
 {\bf \rm Model\  I}:\ \  \mu = 300~{\rm GeV}, \ M_{\rm gaugino} = 900~{\rm
  GeV}, \ M_{\rm mess} = 10^{10}~{\rm GeV}, \ N_{\rm mess} = 1\ ,
\end{eqnarray}
\begin{eqnarray}
 {\bf \rm Model\  II}:\ \  \mu = 500~{\rm GeV}, \ M_{\rm gaugino} = 900~{\rm
  GeV}, \ M_{\rm mess} = 10^{8}~{\rm GeV}, \ N_{\rm mess} = 1\ .
\end{eqnarray}
The $\tilde \tau$ and neutralino masses in Model I are $m_{\tilde \tau}
= 116~{\rm GeV}$ and ($m_{\chi_1^0}$, $m_{\chi_2^0}$, $m_{\chi_3^0}$,
$m_{\chi_4^0})$ = (187~GeV, 276~GeV, 306~GeV, 404~GeV). In Model II,
$m_{\tilde \tau} = 157~{\rm GeV}$ and ($m_{\chi_1^0}$, $m_{\chi_2^0}$,
$m_{\chi_3^0}$, $m_{\chi_4^0})$ = (194~GeV, 346~GeV, 505~GeV,
525~GeV). We chose a parameter with heavier Higgsinos in Model II.

With the SUSY spectra, we have generated 40,000 SUSY events for each
model by using the Herwig~6.50 event generator~\cite{Corcella:2002jc}
with the CTEQ5L parton distribution function~\cite{Lai:1999wy}. This
corresponds to an integrated luminosity of 33~{$\rm fb^{-1}$}
(46~fb$^{-1}$) at the LHC for Model I (Model II).
We set the mixing angle of the right-handed sleptons to be $\sin
\theta_{23} = 0.33$ with which the branching ratio of the lightest
neutralino is $\Gamma (\chi^0_1 \to \tilde \tau \mu) / \Gamma ( \chi^0_1
\to \tilde \tau \tau) \simeq 0.1$. Heavier neutralinos do not have
significant branching ratios for the $\chi^0 \to \tilde \tau \mu$ decays
because the amplitudes are suppressed by the Yukawa coupling constant of
the muon.
With SUSY spectra with the lightest neutralino being almost the Bino and
the lighter $\tilde \tau$ to be almost right-handed (which is the case in the
above two models), the following method is not sensitive to mixings in
left-handed sleptons.
The events are passed through a detector simulator
AcerDET~1.0~\cite{RichterWas:2002ch} where muon momenta are smeared
according to the resolutions of the ATLAS detector. We have also smeared
the momenta and velocities of $\tilde \tau$'s according to the
resolution obtained in Ref.~\cite{stau};
\begin{eqnarray}
 {\sigma(p) \over p} = k_1 p \oplus {k_2 \sqrt{1+{m^2 \over p^2}}}
\oplus {k_3 \over p}\ ,
\end{eqnarray}
where $k_1 = 0.0118\%$, $k_2 = 2\%$ and $k_3 = 89\%$. The momentum $p$
is in GeV. The resolution of the velocity is
\begin{eqnarray}
 {\sigma(\beta) \over \beta} = 2.8\% \times \beta\ .
\end{eqnarray}
We have ignored the $\eta$ dependence of the resolutions. Also, in the
following analysis, we assume that the $\tilde \tau$ mass is known with
a good accuracy by the method of Ref.~\cite{Ambrosanio:2000ik}.

We have followed the strategy of Ref.~\cite{Ambrosanio:2000ik} for the
identification of $\tilde \tau$. We require the candidate tracks to be
within $|\eta| < 2.4$, $P_T > 20$~GeV and $\beta \gamma_{\rm meas} >
0.4$. The cut on the measured velocity ensures $\tilde \tau$ to reach
the muon system.
A consistency condition: $|\beta^\prime - \beta_{\rm meas}| < 0.05$ is
imposed, where $\beta^\prime$ is a velocity calculated from the momentum
($\beta^\prime = \sqrt{p^2/(p^2+m_{\tilde \tau}^2)}$). 
By also requiring the measured velocities of at least one candidate
$\tilde \tau$ to be within $0.4 < \beta \gamma_{\rm meas} < 2.2$ for
each event, this selection strategy reduces background from
mis-identified muons to a negligible level~\cite{Ambrosanio:2000ik}. We
therefore ignore in the following analysis the background from the
standard model processes as well as from muons in SUSY events.

In order to look for lepton flavor violating neutralino decays, we
selected events with only one isolated muon with $P_T > 20$~GeV and at
least one opposite-sign $\tilde \tau$ candidate. If there are two
opposite-sign $\tilde \tau$-$\mu$ pairs, we use both of them for the
analysis. The invariant mass $M_{\tilde \tau \mu}$ is calculated for
each candidate event.

\begin{figure}[t]
\begin{center}
  \includegraphics[height=6.5cm]{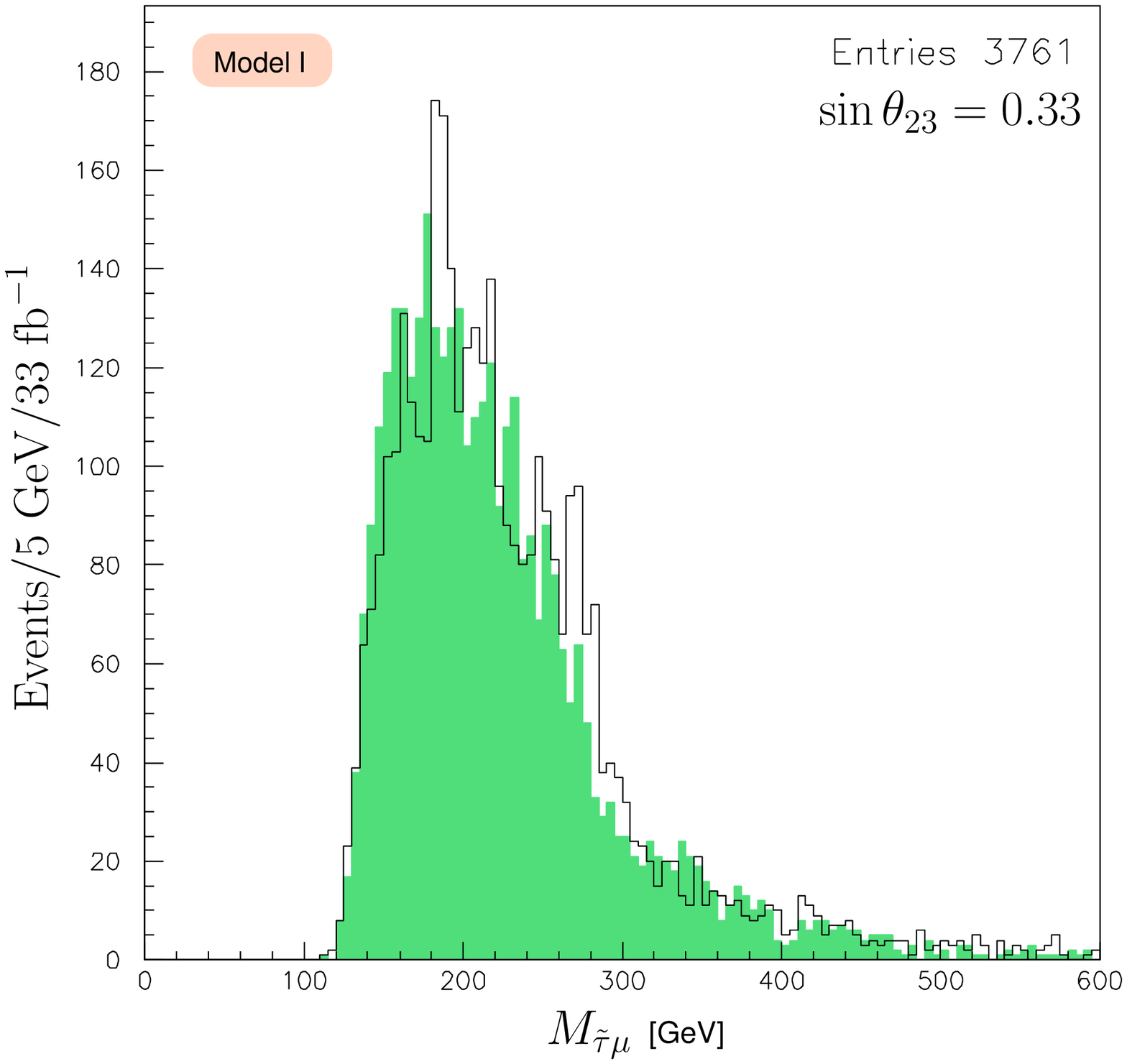}\hspace*{3mm}
  \includegraphics[height=6.5cm]{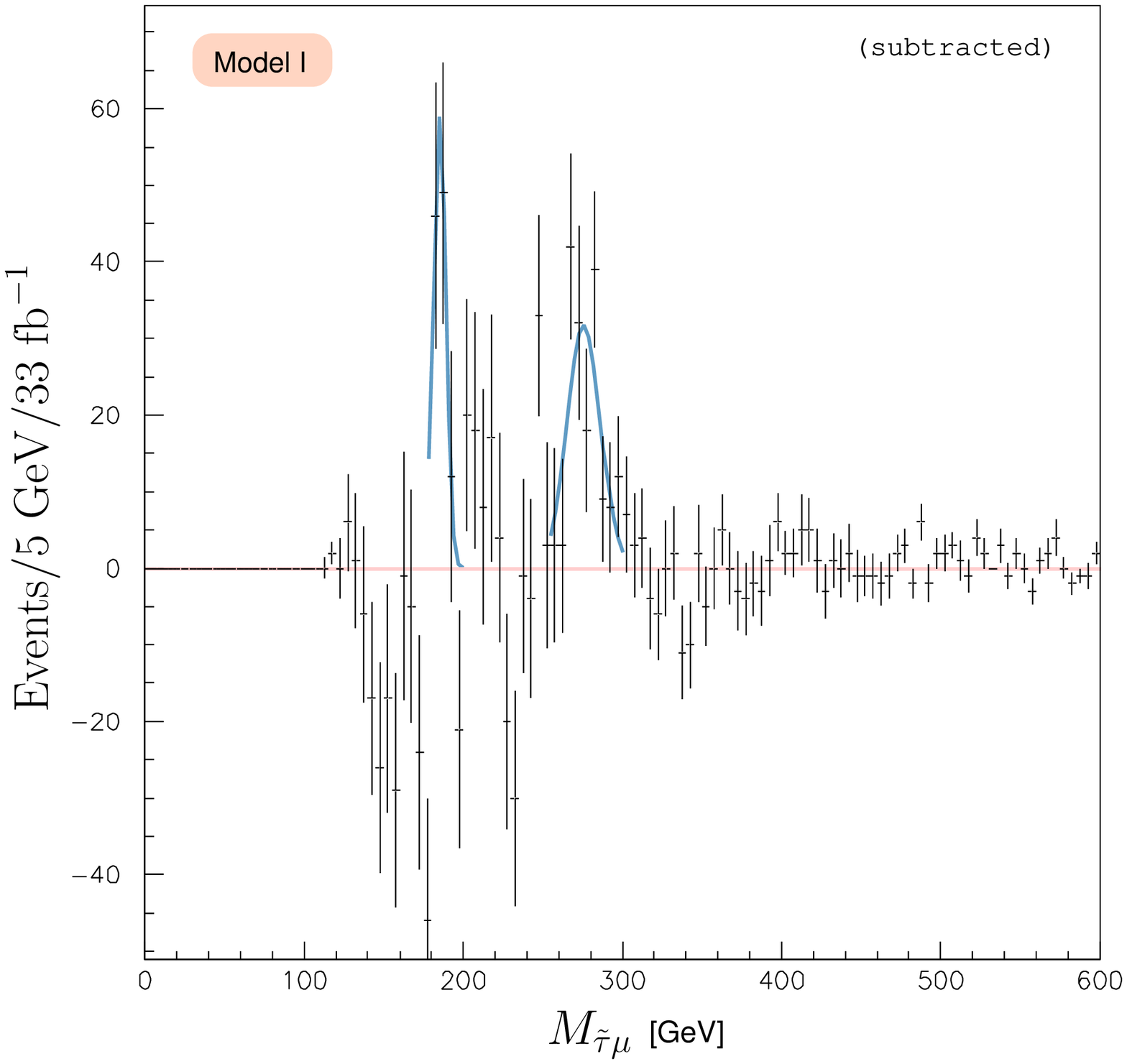}
\end{center}
\caption{The $\tilde \tau$-$\mu$ invariant mass distribution in Model
 I. The shaded histogram in the left panel is the estimated background,
 i.e., the $\tilde \tau$-$e$ invariant mass distribution. By subtracting
 the estimated background, we obtain the histogram in the right panel.}
\label{fig:model-1}
\end{figure}

The invariant mass distribution is shown in the left panel of
Fig.~\ref{fig:model-1}, where we can clearly see a peak at the lightest
neutralino mass (187~GeV). The shape and normalization of the background
distribution can be obtained from the $M_{\tilde \tau e}$ distribution
directly from the data (shaded histogram). By subtracting those
estimated background, we obtain the histogram in the right panel where
we see that the background is successfully subtracted. Therefore we can
reliably use the $M_{\tilde \tau e }$ distribution as an expected
background. We can also find an excess around masses of the lighter
Higgsino-like neutralino ($\sim 276$~GeV).
Most of the background originates from the $\chi^0 \to \tilde \tau \tau$
decay followed by $\tau \to \mu \nu \bar \nu$. Although the signal
region is the kinematic endpoint of this background for each neutralino,
the background $\tilde \tau \mu$ pairs from heavier neutralinos fall
into the signal region. There are also backgrounds from leptonic decays
of $W$ bosons. Numbers of such background events depend on cascading
pattern of heavy SUSY particles.

We fitted two peaks in the right panel of Fig.~\ref{fig:model-1} with
the gaussian function and defined the signal region to be the 1$\sigma$
region around the peaks; $|M_{\tilde \tau \mu} -185.2~{\rm GeV}
|<3.8~{\rm GeV} $ and
$|M_{\tilde \tau \mu} -276~{\rm GeV} |<10~{\rm GeV} $.
In the case where there are not enough events
to find the neutralino masses by the $M_{\tilde \tau \mu}$ distribution,
one should look for edges in the invariant mass of $\tilde \tau$ and
$\tau$-jet, $M_{\tilde \tau j_\tau}$, for the neutralino mass
measurements as is done in Ref.~\cite{Ibe:2007km}.
There are $S+B=584$ events in the signal region whereas the number of
the expected background in the signal region is $B=374$. Therefore we
obtain $9\sigma$ excess with 33~fb$^{-1}$ of data.\footnote{This level
of excess is somewhat optimistic given that we know the correct location
of the peaks. In the actual experimental situation, the peak locations
(the neutralino masses) will be measured by looking for the endpoint
locations of the invariant mass $M_{\tilde \tau j_\tau}$. The
uncertainty of this measurement is estimated to be at most of order
$5\%$~\cite{Ibe:2007km} by taking into account the effects of fake
$\tau$-jets and the uncertainties in calibration of the $\tau$-jet
energies. If we use the central values given in Table 2 of
Ref.~\cite{Ibe:2007km} and the $5\%$ errors for the definition of the
signal region, i.e., $194 \pm 10$~GeV and $279 \pm 14$~GeV, we obtain
about a $7\sigma$ excess, where most of the significance is a
contribution from the second peak.} Normalizing to the integrated
luminosity of 100~fb$^{-1}$, the number of needed signal events for
5$\sigma$ discovery to be 181, corresponding to $\sin \theta_{23} >
0.18$.

\begin{figure}[t]
\begin{center}
  \includegraphics[height=6.5cm]{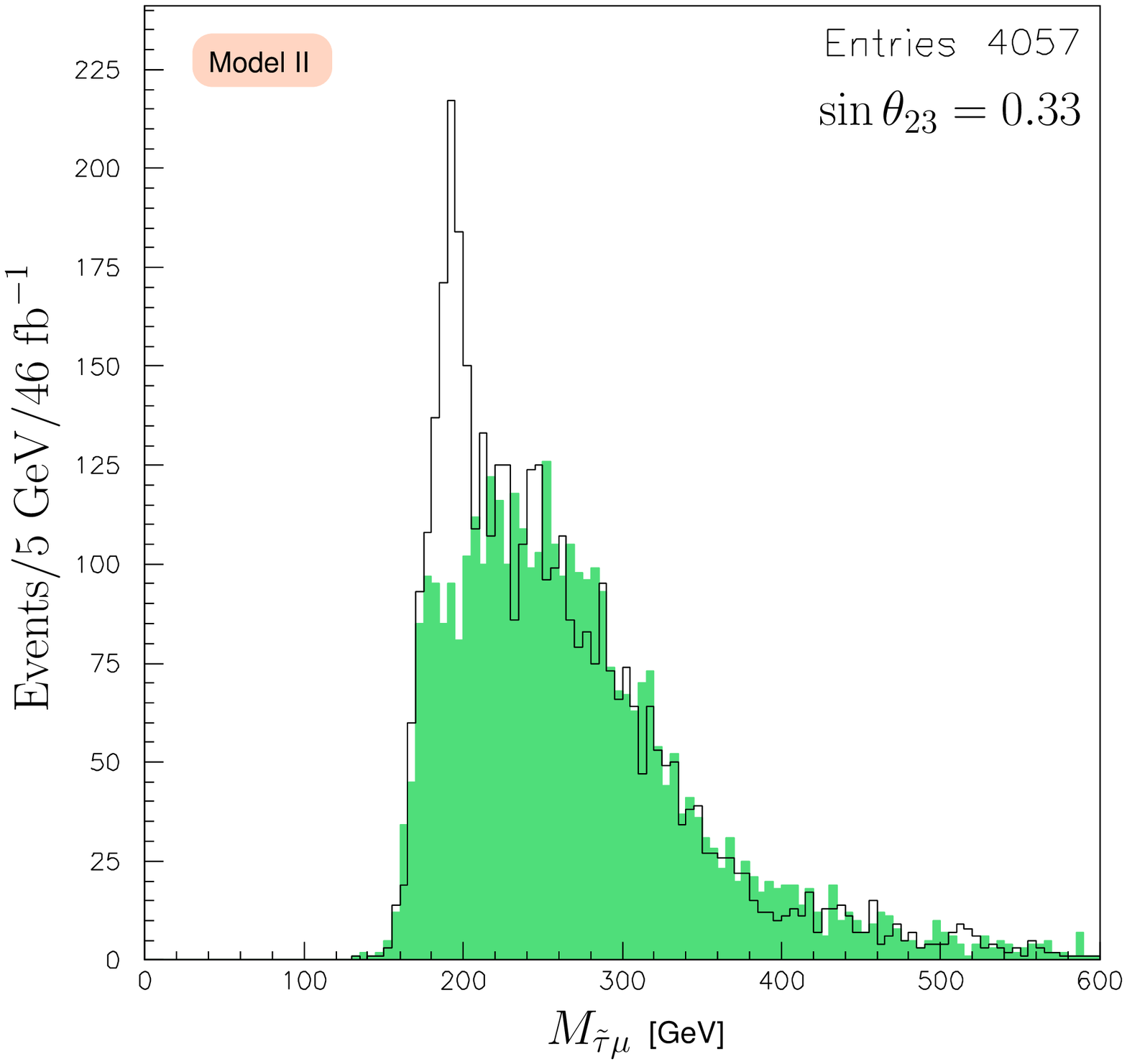}\hspace*{3mm}
  \includegraphics[height=6.5cm]{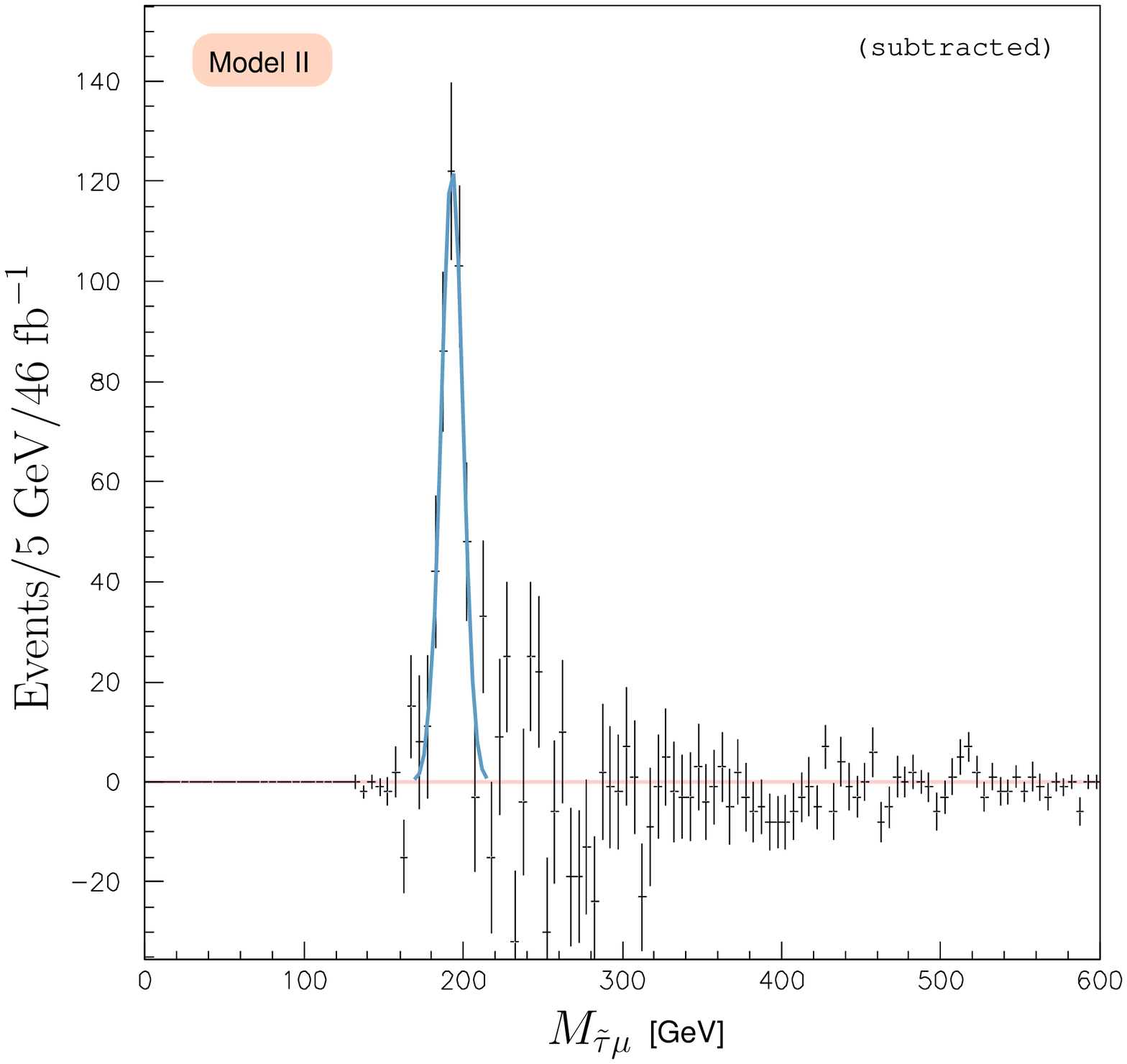}
\end{center}
\caption{The $\tilde \tau$-$\mu$ invariant mass distribution in Model
 II. The shaded histogram in the left panel is the estimated background,
 i.e., the $\tilde \tau$-$e$ invariant mass distribution. By subtracting
 the estimated background, we obtain the histogram in the right panel.}
\label{fig:model-2}
\end{figure}

We repeat the same analysis for the Model II and the result is shown in
Fig.~\ref{fig:model-2}. There is only one peak associated with the
Bino-like neutralino because the number of Higgsinos in cascade decays
is reduced and the branching ratios of $B(\chi^0_{3,4} \to \tilde \tau
\tau)$ are suppressed with the relatively heavy Higgsinos.
A slightly better sensitivity than Model I is obtained. In the signal
region, $|M_{\tilde \tau \mu} -193.0~{\rm GeV} |<6.9~{\rm GeV} $, we
find $S+B=539$ and $B=238$ for 46~fb$^{-1}$. Normalizing to
100~fb$^{-1}$ of data, we obtain the 5$\sigma$ sensitivity to be $\sin
\theta_{23} > 0.15$.

If no peak is found due to small mixing angles, one can put a bound on
the branching fraction (or equivalently the mixing angle). This requires
a counting of the $\chi^0 \to \tilde \tau \tau$ events that involves the
efficiency measurement of the $\tau$ identification. That will be the
dominant uncertainty in putting the experimental bound. As far as order
of magnitude is concerned, the sensitivity will be at the level of $\sin
\theta_{23} \sim 0.1-0.2$.

The search for the $\tau \to \mu \gamma$ decay has already put a
stringent bound on a combination of various SUSY parameters involving
the slepton mixings. Although a model independent comparison is not
possible, we can get a sense of sensitivities to the mixing parameter by
calculating the $\tau \to \mu \gamma$ branching ratio with a particular
parameter set. We have done that in Model I and II with $\sin
\theta_{23} = 0.15$. The branching ratios are $B(\tau \to \mu \gamma) = 1
\times 10^{-6}$ and $4 \times 10^{-9}$ for Model I and II,
respectively. Compared with the current experimental bound, $4.5 \times
10^{-8}$~\cite{Hayasaka:2007vc}, the LHC sensitivities can be much
better (or worse) depending on model parameters. One should also note
that the measurement of $\Gamma(\chi^0 \to \tilde \tau
\mu)/\Gamma(\chi^0 \to \tilde \tau \tau)$ at the LHC will directly probe
the slepton mixing parameter. Therefore, measuring/constraining the
branching fractions of both processes will be important to understand
the flavor structure of SUSY models. We summarize the results in
Table~\ref{tab}.

\begin{table}[t]
\begin{center}
 \begin{tabular}{ccccc||cc}
\hline \hline
  & $\sigma_{\rm SUSY}$ (${\cal L}$) & $S+B$ & $B$ & $S / \sqrt{S+B}$ &
  $\sin \theta_{23}^{\rm min}$ (100~fb$^{-1}$) & $B(\tau \to \mu
  \gamma)$ \\ \hline 
  Model I & 1.2~pb (33 fb$^{-1}$)& 584  & 374 & 9 & 0.18 & $1\times 10^{-6}$ \\ \hline
  Model II& 0.88~pb (46 fb$^{-1}$)& 539  & 238 & 13 & 0.15 & $4\times 10^{-9}$\\ \hline \hline
 \end{tabular}
\end{center}
\caption{LHC sensitivities and comparison to the $\tau \to \mu \gamma$
decay. The number of signal $(S)$ and background $(B)$ events are shown
for 40,000 SUSY events and $\sin \theta_{23}= 0.33$. For this level of
the large mixing angle, the statistical significances can be as large as
10$\sigma$. The 5$\sigma$-level discovery with an integrated luminosity
of $100$~fb$^{-1}$ requires the angle to be $\sin \theta_{23} >
0.15$. The $\tau \to \mu \gamma$ branching ratios are shown for $\sin
\theta_{23} = 0.15$.}  \label{tab}
\end{table}

A similar analysis will go through for a $\chi^0 \to \tilde \tau e$
search at the LHC. Also, if a linear collider is built in future,
searches for a decay mode $\chi^0 \to \tilde \tau \mu$ through a
neutralino pair production process may give a better sensitivity to the
mixing angle as background from heavier neutralinos and $W$ bosons will
be under better control.

\section*{Acknowledgments}

I thank Alex Friedland for discussion and useful comments.


\end{document}